\documentstyle[aps]{revtex}

\def\be{\begin{equation}}
\def\ee{\end{equation}}
\def\ba{\begin{eqnarray}}
\def\ea{\end{eqnarray}}

\def\A{{\cal A}}

\def\Ab{{\overline{\A}}}
\def\Rl{{\mathchoice
	{\setbox0=\hbox{$\displaystyle\rm R$}\hbox{\hbox to0pt
	{\kern0.4\wd0\vrule height0.9\ht0\hss}\box0}}
	{\setbox0=\hbox{$\textstyle\rm R$}\hbox{\hbox to0pt
	{\kern0.4\wd0\vrule height0.9\ht0\hss}\box0}}
	{\setbox0=\hbox{$\scriptstyle\rm R$}\hbox{\hbox to0pt
	{\kern0.4\wd0\vrule height0.9\ht0\hss}\box0}}
	{\setbox0=\hbox{$\scriptscriptstyle\rm R$}\hbox{\hbox to0pt
	{\kern0.4\wd0\vrule height0.9\ht0\hss}\box0}}}}

\begin{document}

\twocolumn[\hsize\textwidth\columnwidth\hsize\csname
@twocolumnfalse\endcsname

\title{The Immirzi parameter in quantum general relativity}

\author{Carlo Rovelli\footnote{rovelli@pitt.edu}}
\address{Physics Department, University of Pittsburgh,
Pittsburgh, PA 15260, USA\\
Center for Gravity and Geometry, Penn State University, USA}

\author{Thomas Thiemann\footnote{thiemann@math.harvard.edu}}
\address{Physics Department, Harvard University, Cambridge, 
MA 02118, USA}

\date{\today}

\maketitle

\begin{abstract}
Barbero has generalized the Ashtekar canonical transformation to  
a one-parameter scale transformation $U(\iota)$ on the phase space of 
general relativity.  Immirzi has noticed that in loop quantum 
gravity this transformation alters the spectra of geometrical
quantities.  We show that $U(\iota)$ is a canonical transformation that 
cannot be implement unitarily in the quantum theory.  This implies that 
there exists a one-parameter quantization ambiguity in quantum gravity, 
namely a free parameter that enters the construction of the quantum theory.
The purpose of this letter is to elucidate the origin and the role of 
this free parameter.

Preprint HUTMP-97/B-366 \hskip 7.4cm     PACS: 04.60.Ds

\end{abstract}
\vskip.5cm
]

\section{Introduction}

One of the most interesting recent results in quantum gravity is the 
computation of the quantum spectra of certain geometrical quantities 
\cite{spectra}.  These spectra turn out to be discrete, suggesting a 
discrete structure of geometry at the Planck scale which might be, in 
principle, physically measurable \cite{disc}.  Recently, Immirzi has 
pointed out that the overall scale of these spectra is not determined 
by the theory \cite{Immirzi}.  More precisely, Immirzi has considered a 
certain scale transformation introduced by Barbero \cite{Barbero}, and 
noticed that if we quantize the theory starting from {\em scaled\/} 
elementary variables, we end up with different spectra for the 
same geometrical quantities.

Here, we analyze the issue in detail.  We find that the quantum theory 
is in fact undetermined by one para\-me\-ter.  This is due to the fact 
that the holonomy algebra on which this approach is based depends on 
a free parameter.  This fact gives rise to a one-parameter family of 
inequivalent quantum theories, which are all, up to additional physical 
inputs, physically viable. In a sense, there is a one-parameter 
family of ``vacua'' in quantum general relativity, parameterized by a 
free (real) parameter, which we call ``Immirzi parameter'', and denote 
as $\iota$ (``iota'').  Equivalently, there is a symmetry in the 
classical theory which is realized as a canonical transformation but 
cannot be realized as a unitary transformation in the quantum theory.

The existence of this quantization ambiguity is due to the peculiar 
kind of representation on which non-perturbative quantum gravity is 
based \cite{loops,AI}.  This representation is characterized by the 
fact that the holonomy is a well-defined operator in the quantum 
theory.  Conventional perturbative Maxwell and Yang-Mills theories are 
not defined using this kind of representation and the $\iota$ 
parameter does not appear in that context.  But physical and 
mathematical arguments indicate that this representation is relevant at the 
diffeomorphism-invariant and background-independent level 
\cite{AI,ascona}.  Thus, the Immirzi parameter appears in the general 
covariant context.

In this letter, we describe in some detail how this ambiguity is 
originated and its consequences.  In particular, we address a certain
number of questions that have been recently posed concerning the $\iota$
parameter, and we try to rectify a number of proposed incorrect 
interpretations of the appearance of this free parameter.

For a similar discussion, but centered on the loop quantization of Maxwell 
theory, see \cite{k}, where an interesting speculation tying the Immirzi 
parameter with charge quantization is presented.

\section{The Immirzi ambiguity}

Consider a three-dimensional compact smooth manifold $\Sigma$, and two 
fields defined over it: an $su(2)$ valued vector density $E$, with 
components $E^a_i$, where $a,b,c,..$ denote tensor indices, and 
$i,j,k,..$ are $su(2)$ indices; and an $SU(2)$ connection $A$ with 
components $A_a^i$.  In terms of these fields, one can describe Euclidean 
\cite{Abhay} or Lorentzian \cite{Barbero} general 
relativity, as well as $SU(2)$ Yang-Mills theory.  In both cases $A_a^i$ 
and $E^a_i$ are canonically conjugate and subject to the Gauss constraint.
\be \label{Gauss}
            G_i:=\partial_a E^a_i+\epsilon_{ijk }A^j_aE^{ak}.
\ee
In Yang-Mills theory, $E$ is the electric field; in general relativity,
$E^a_i=\det(e_a^i)\,e^a_i$, were $e_a^i$ and $e^a_i$ are the triad field
(related to the 3-metric by $g_{ab}=e_a^ie_b^i$) and its inverse. 

In general relativity it is natural to take dimensionless coordinates. 
$E$ has dimension $length^2$ and $`A=A_a dx^a$ is dimensionless.  The 
fundamental Poisson brackets are $\{A_a^i(x),E^b_j(y)\}=G\, \delta_a^b 
\delta^i_j \delta(x,y)$, where $G$ is $16\pi\,c^{-3}$ times the Newton 
constant.

We denote by $h_e(A)\in SU(2)$ the transport propagator of $A$ along a path 
$e$.  If $e$ is a closed path then $\ \mbox{tr}(h_e(A))$ is the holonomy 
of $A$ around $e$, or the Wilson loop functional.  Wilson loops are 
natural $SU(2)$ gauge invariant quantities, and any gauge invariant 
function of $A$ can be approximated by sum of products of Wilson loops. 
In conventional perturbative Yang-Mills theory, there is no 
well-defined operator corresponding to the Wilson loop, because field 
operators need to be smeared in more than one dimension to be well 
defined. In a general covariant theory, on the other hand, one expects the 
Wilson loop operators to be well-defined in the quantum theory.   
Thus, one can define the quantum theory as a unitary representation of the 
(non-canonical) Poisson algebra generated by the classical observables 
$h_e(A)$ and $E^a_i(x)$ (with a suitable smearing of the later) 
\cite{loops}. 

By doing so, one obtains a quantum representation that can be 
described as follows \cite{ascona,AL1,ALMMT-JMP}.  The Hilbert space 
of the theory is taken as ${\cal H}= L_2(\Ab,d\mu_0)$, where $\Ab$ is 
(the closure in a suitable norm of) the space of the smooth connection 
fields, and $d\mu_0$ is a faithful, diffeomorphism invariant 
$\sigma$-additive probability measure on $\Ab$.  One can show 
\cite{ALMMT-JMP} that the measure $\mu_0$ is the unique probability 
measure on $\Ab$ which implements the classical reality conditions 
($A_a^i,E^a_i$ are real) and the given Poisson algebra.  The holonomy 
observable $h_e(A)$ becomes the quantum operator $\hat h_e$ defined by
\be 
       \hat h_e \Psi = h_e \Psi, \ \ \ \ \Psi\in{\cal H}.  
\ee

A characteristic example of a geometric operator defined in the theory 
is the area of a two-dimensional surface.  Recall that 
the area $A(S)$ of a 2d surface $S$ immersed in $\Sigma$ is given by 
\be
   A(S)=f(S,E)
\label{area}
\ee
where
\be
     f(S,E) = \int_S \sqrt{n_a\, n_b\, E^a_i\, E^{bi}}\ \ ;
\ee
$n_a$ is the one-form normal to the surface $S$. For every given surface 
$S$, the quantity $f(S,E)$ yields (via a quantization and regularization 
procedure) a well-defined operator $\hat f(S)$ on $\cal H$.  Its 
spectrum is discrete and positive \cite{spectra}. For more details, 
see \cite{ascona}. 

Now, the triad $e_a^i$ defines a three dimensional 
spin connection $\Gamma_a^i$ by
\be
  \partial_{[a} e_{b]}^i = \epsilon^{i}{}_{jk} \Gamma_{[a}^j e_{b]}^k,
\label{gamma}
\ee
which transforms under $SU(2)$ as the connection $A$ does. 
Pick an arbitrary positive real number $\iota$ and define the scaled 
fields
\begin{eqnarray}
A_\iota{}_a^i &=& \iota A_a^i+(1-\iota)\Gamma_a^i, \label{atransf} \\
E_\iota{}^a_i &=& {1\over\iota}E^a_i.
\end{eqnarray}
It is then easy to verify the following. i) $A_\iota$ and $E_\iota$ are 
canonically conjugate, namely the map $U_\iota:(A,E) \mapsto 
(A_\iota,E_\iota)$ is a canonical transformation.  ii) $A_\iota$ and 
$E_\iota$ transform under gauge transformations as $A$ and $E$ do.  In 
particular, $A_\iota$ is a connection.  This follows from the fact 
that $A_\iota$ is a convex linear combination of two $SU(2)$ 
connections.  Thus $U_\iota$ preserves the affine structure of the 
space of connections.  iii) Clearly $U_\iota$ preserves the reality 
conditions as well.  Thus $U_\iota$ leaves the kinematical structure 
of general relativity invariant.

The canonical transformation $U(\iota)$ was studied by Barbero 
\cite{Barbero}.  The case $\iota=\sqrt{-1}$ corresponds to 
Ashtekar's canonical transformation \cite{AbhayBeta}.  Thus, the fact 
that $U(\iota)$ is a canonical transformation is a simple extension 
of the important discovery due to Ashtekar \cite{Abhay} which gave 
rise to the connection formulation of general relativity.  To see the 
relation with these works, notice that by introducing the extrinsic 
curvature $K_a^i=A_a^i-\Gamma_a^i$, we can write (\ref{atransf}) as
\begin{eqnarray}
A_a^i &=& \Gamma_a^i + K_a^i, \\
A_\iota{}_a^i &=& \Gamma_a^i + \iota K_a^i.
\end{eqnarray}
The generating function of the (infinitesimal) canonical transformation 
is $C=\int K_a^iE_i^a$. Finally, notice that the area of a surface $S$ is 
given in terms of the scaled variables by
\be
   A(S) = \iota \ f(S,E_\iota).
\label{scaledarea}
\ee

Now, we can view $(A_\iota,E_\iota)$ just as different coordinates on 
the phase space of general relativity.  Suppose we decide to quantize the 
theory following the same path as before, but starting from the scaled 
variables  $(A_\iota,E_\iota)$.  In general, starting a quantization 
from scaled variables has no effect on the physical predictions of the 
quantum theory.  For instance, we may quantize a harmonic oscillator 
starting from scaled canonical coordinates $({1\over\iota}q,\iota p)$ 
without affecting, say, the spectrum of the hamiltonian -- in the next 
section we shall verify that this is indeed the case.  Similarly, it 
is natural to expect here that physical predictions are not going to 
be affected if we quantize the variables $(A_\iota,E_\iota)$ instead 
of the variables $(A,E)$.  More precisely, we expect that the 
canonical scale transformation $U(\iota)$ will be implemented 
unitarily in the quantum theory, so that the quantization based on 
$(A_\iota,E_\iota)$ will yield a different, but unitary equivalent 
quantum representation.  However, this expectation turns out to be 
false.

Indeed, let us define the theory requiring that the pa\-rallel transport 
$h_e^\iota=h_e(A_\iota)$ be a well defined operator in the quantum 
theory.  Call ${\cal H}_\iota$ the Hilbert space on which we define 
the new representation.  Since the Poisson algebra formed by the 
quantities $h_e^\iota$ is the same as the one formed by the quantities 
$h_e$, we may use the same Hilbert space as before: ${\cal H}_\iota = 
L_2(\Ab,d\mu_0)$.  (We denote the two Hilbert spaces $\cal H$ and 
${\cal H}_\iota$ differently because the interpretation of the same 
function $\Psi\in L_2(\Ab,d\mu_0)$ is different in $\cal H$ and in ${\cal 
H}_\iota$).

If the two quantization paths define the same theory, there should 
exist a unitary transformation $U(\iota)$ from $\cal H$ to ${\cal H}_\iota$ 
which sends physically corresponding states and physically corresponding 
operators into each other.  Namely, we expect that the canonical 
transformation $U(\iota)$ can be realized as a unitary transformation of 
$L_2[\bar{\cal A},d\mu]$ into itself.  

If this transformation exists\footnote{The generator of the classical 
canonical transformation can be promoted to a self-adjoint operator. 
It coincides with the generator of the Wick rotation transform 
considered in \cite{Wick}. The corresponding operator $\hat{C}$ 
is defined on $\cal H$ in \cite{QSD}}, an immediate consequence is 
that the spectra of observables which are unitary images of each other 
are identical.  In particular, the spectrum of the area operator 
defined in one theory should agree with the spectrum of the area 
operator defined in the other theory.  

Consider now the operator $f(S,E_\iota)$ defined in the scaled theory.  
Since it is has the same form as the operator $f(S,E)$ in the unscaled 
theory, it will certainly have the same spectrum.  But then equations 
(\ref{area}) and (\ref{scaledarea}) show that the spectrum of the area 
in the scaled theory is obtained from the one in the original theory by 
multiplying it by $\iota$.  Since the spectrum of the area is 
discrete, it cannot be invariant under a scaling by an arbitrary real 
parameter.  Therefore the two quantizations yield unitarily 
inequivalent theories.

The two theories give different physics. For instance, the predicted 
spectrum of the area differ.  The other geometrical quantities having 
discrete spectrum, such as volume and length, have different spectra 
in the scaled theory as well.  Since $\iota$ enters linearly in $E$, 
which has dimension $L^2$, the discrete spectrum of a geometrical 
quantity homogeneous in $E$ with dimension $L^n$ scales as 
$\iota^{n/2}$.

Thus, there is a free parameter $\iota$ in the quantization of the 
theory.  It could be measured, in principle, simply by measuring, for 
instance, the size of the ``quanta'' of area.  So far, there does not 
seem to be any compelling reason for choosing a particular value for 
$\iota$.  Better knowledge of the theory may indicate such a reason.

It has been suggested (for instance, see \cite{k}) that one could fix 
$\iota$ using black hole entropy.  It was shown in \cite{blackhole} 
that one can derive the Bekenstein-Hawking entropy formula Entropy 
$ = const \times A$ from loop quantum gravity -- with a constant $const$ 
that turns out to be finite, but with an incorrect value.  
This constant is affected by a rescaling of $\iota$, and therefore one 
might choose a value for $\iota$ yielding the correct value 
$const={1\over 4}\hbar G$.  This would fix a value of $\iota$ 
approximately given by 
\be 
    \iota \sim {1\over 4 \pi}.  
\ee

\section{Incorrect interpretations}

Several interpretations of the appearance of the free parameter 
$\iota$ in quantum gravity have been recently proposed.  Some 
of these are incorrect.  Here we discuss some of these 
interpretations.

\subsection{$\iota$ is a free constant that multiplies the connection in 
the definition of the holonomy}\label{ssholo}

Let $e:[0,2\pi]\mapsto\Sigma$ be a closed path. The holonomy $h_e$ is 
formally written as 
\be
	h_e = {\cal P} e^{\oint_e A},
\label{exp}
\ee
and it is more precisely defined as $h_e = h_e(2\pi)$ where the 
$SU(2)$-valued function $h_{e}(s)$ is the solution of the differential 
equation
\be 
{d\over ds}h_e(s) - \dot e^a(s) A_a(e(s))\ h_e(s)=0,\ \ \ \ h_e(0)=1.
\ee 
Here $\dot e^a(s)$ is the tangent to $e$ at $s$ and $1$ the identity 
in $SU(2)$.  It has been repeatedly suggested that in defining the 
holonomy one if free to add a dimensionless parameter $\iota$ in the 
exponent of (\ref{exp}) -- keeping any other thing fixed in the 
theory.  

However, this is not true.  The transformation properties of $A$ under 
an $SU(2)$ gauge transformation are uniquely fixed by the classical 
action, because from the action one derives the gauge constraint 
(\ref{Gauss}).  If (\ref{exp}) is gauge covariant, then \be \tilde h_e 
= {\cal P} e^{\iota\oint_e A},
\label{exp2}
\ee
or, more precisely, the solution of 
\be
{d\over ds}\tilde h_e(s) - \iota \dot e^a(s) A_a(e(s))\ \tilde h_e(s)=0,
 \ \ \tilde h_e(0)=1,
\ee
is {\em not\/} gauge covariant, as a direct computation shows. In other 
words, if we multiply a connection by a real number, we do not obtain a 
quantity that transforms as a connection: connections form an affine 
space, not a linear space. 

\subsection{$\iota$ is the constant that multiplies the classical action}
\label{ssactio}

The quantity $\tilde h_e$ defined in (\ref{exp2}) is gauge covariant 
under the gauge transformation generated by a scaled Gauss constraint
\be 
\label{Gauss2}
        G_i:=\partial_a E^a_i + \iota \epsilon_{ijk }A^j_aE^{ak}. 
\ee 
This constraint, in turn, can be obtained by scaling the action of the 
theory.  Accordingly, it has been suggested that the $\iota$ 
ambiguity is a consequence of the freedom in choosing the constant in 
front of the action of general relativity. Equivalently: scaling 
$A$, but not $E$, without absorbing this into a redefinition of the 
coupling constant.

Again, this interpretation is wrong. Physically, the constant in front of 
the general relativity action determines the strength of the macroscopic 
Newtonian interaction.  The freedom in the choice of the Immirzi parameter 
in the quantum theory consists in the fact that the overall scale of the 
spectra is {\em not\/} determined by low energy physics.  In other 
words, we can measure the Newton constant by means of {\em 
classical\/} gravitational experiments, and measure the Planck 
constant by means of {\em non-gravitational\/} quantum experiments.  
From these two quantities we obtain a length, the Planck length 
$l_{P}=\sqrt{\hbar G}$.  The point of the Immirzi ambiguity is 
that the ratio of, say, a given eigenvalue of the area to $l_P$ is not 
determined by the quantization procedure.

\subsection{Any Yang-Mills theory has a free $\iota$ parameter when 
quantized in the loop representation}\label{ssany}

As the previous comments indicate, the Immirzi parameter is not 
related to a multiplicative scaling of the connection.  Rather, it is 
generated by the affine transformation (\ref{atransf}).  In order to 
be able to write such an affine transformation, we need to have a 
second, independent connection in the formalism.  In general 
relativity, where the hamiltonian gauge group is $SU(2)$, the 
``electric field'' $E^{a}_i$ has the correct index structure for a 
connection $\Gamma_a^i$ to be defined via (\ref{gamma}).  
(Furthermore, $\Gamma_a^i$ has a natural geometrical interpretation.)  
There is no direct analog of this construction, as far as we can see, 
for an arbitrary Yang Mills theory.\footnote{$SU(2)$ Yang-Mills theory 
has essentially the same phase space as general relativity.  However, 
notice that in order to define $\Gamma_a^i$ we need to invert the 3x3 
matrix $E^{a}_{i}$; in general relativity this matrix is invertible, 
because of the non-degeneracy condition on the metric, but in $SU(2)$ 
Yang-Mills it is not so in general.}

An exception of the above is the case of an abelian theory, namely for 
Maxwell theory.  In the case of Maxwell, the quantity (\ref{exp2}) is 
gauge invariant for any $\iota$.  Thus, we have a free parameter in 
the loop representations of electromagnetism and general relativity 
only.  In the ($U(1)$) Maxwell theory, the free parameter appears 
because a scaled abelian connection is still a connection.  In the 
($SU(2)$) general relativity it appears because a second connection 
for defining an affine transformation is determined by the 
triad.  There is no obvious generalization of this effect to other 
gauge theories.

\subsection{The algebra has inequivalent representations}

A characteristic phenomenon in field theory --and in finite dimensional 
quantum mechanics, when the phase space is non linear-- is that the main 
observable algebra may have inequivalent representations. It has 
been suggested that the Immirzi parameter distinguishes inequivalent 
representations of the same algebra. However, we do not think that this is 
the case.  We have defined the $\iota$-scaled theory as a 
representation of a physically distinct algebra (the algebra of the 
scaled variables) obtaining the quantum theory on ${\cal H}_\iota$.  
If this theory could be obtained also as a representation of the 
original algebra of unscaled variables, then it would carry a 
representation of the scaled as well as the unscaled, $h_e$ and 
$h_e^\iota$, operators.  But we do not see any reason for the 
operators $h_e^\iota$ to be well-defined in the Hilbert space 
of the usual loop quantum theory.

\section{Models}

We now discuss a few simple models in which the Immirzi ambiguity does 
or does not appear, in order to illustrate some of the statements made 
above.

\subsection{Harmonic oscillator: no $\iota$ ambiguity}

Consider the phase space $\Gamma=\Rl^2$ of an harmonic oscillator with 
canonical coordinate and momentum $(x,p)$ and Hamiltonian 
\be
       H = {1\over 2}p^2+{\omega^2\over 2} x^2.
\ee 
The quantum theory is defined by a representation of the canonical 
algebra $[q,p]=i\hbar$.  Let the Hilbert space be ${\cal H}=L_2[R,dq]$ 
and $(\hat q=x,\hat p=-i\hbar\partial_q)$. The spectrum of the Hamiltonian 
operator is 
\be
	E_n = \left(n+{1\over 2}\right)\hbar\omega.
\ee

Suppose we decide to quantize the theory starting from the scaled variables
\be
     x_\iota = \iota x, \ \ \ \ p_\iota = {1\over\iota} p.  
\ee 
Is the spectrum of the Hamiltonian altered?  The transformation 
$(x,p)\to (x_\iota,p_\iota)$ is clearly a canonical transformation.  
The algebra of the scaled variables is the same as the algebra of 
the unscaled ones.  Therefore we can write ${\cal H}_\iota 
=L_2[R,dq]$ and $(\hat q_\iota=x,\hat p_\iota=-i\hbar\partial_q)$.  
In the scaled variables, the Hamiltonian reads 
\be 
     H = \iota^2 \left[{1\over 2}p_\iota^2+{\omega^2\over 2\iota^4} 
     x_\iota^2\right].  
\ee 
Thus its spectrum is $\iota^2$ times the spectrum of an 
Hamiltonian of angular frequency $\omega/\iota^2$.  That is 
\be 
   E_n = \iota^2 \left(n+{1\over 2}\right)\hbar{\omega\over\iota^2}, 
\ee 
precisely as before.  Therefore, in the quantization based on the 
scaled variables the wave functions $\psi(q)$ have a different 
interpretation, but physical predictions, such as the spectrum 
of the Hamiltonian, are not altered.  This example shows that in 
general a one-parameter scaling of the quantization variables that 
does not alter the observable algebra will {\em not\/} alter the 
theory. This was to be expected from Von-Neumann's uniqueness theorem.

\subsection{Particle on a circle: no $\iota$ ambiguity}

As a second example, we consider another simple system in which a scaling 
of the basic variables does not alter the quantum theory: the theory of a 
particle on a circle. We include this case because, unlike the 
harmonic oscillator case, here the observable algebra has a one-parameter 
family of inequivalent irreducible representations. However, this fact 
turns out to irrelevant as far as the $\iota$ ambiguity is concerned.

Consider a particle constrained to move on a circle of length $L$. Let 
$q\in[0,L]$ its position and $p$ its momentum. The Hamiltonian is 
$H={1\over 2} p^2$, and has eigenvalues $E_n = n^2 {\pi^2\hbar^2\over 
2 L^2}$.  Now, suppose we change basic variable to $ (s_\iota,p_\iota)  = 
(\iota q,{1\over \iota}p)$. Then the Hamiltonian reads $H={\iota^2\over 
2}p^2_\iota$. At first sight one is tempted to say that the spectrum of 
$H$ has now changed, because $p_\iota$ and $q_\iota$ have the same 
algebra as $p$ and $q$ and therefore $p_\iota$ must have the same 
spectrum as $q$.  But of course this is wrong, because $q$ and 
$q_\iota$ are defined on a circle, and the spectrum of their conjugate 
momentum depends on the size of the circle.  Clearly this exactly 
compensates the $\iota^2$ in $H(p_\iota)$.

More precisely, we cannot quantize the $(q,p)$ algebra in this case 
because $q$ is not a global coordinate on the phase space \cite{Isham}. 
We must, for instance, replace $q$ with $g={L\over 2\pi}\exp(i 2\pi q/L)\in 
U(1)$,  which  is a global coordinate. If we introduce also the angular 
momentum $l={L\over 2\pi} p$, we obtain the algebra
\be
	\{g,l\}= g.
\ee
This algebra has a one-parameter family of distinct irreducible  
representations, which can all be defined on $L_2[S_1,d\phi]$ by
\be
  \hat l = -i\partial_\phi, \ \ \ \ \hat g = r e^{i\phi},
\ee
where $r$ labels the representation. The representation 
appropriate for a given $L$ is the one obtained by choosing $r=L/2\pi$, 
because this realizes the non linear condition on the basic variables 
$|g|=L/2\pi$.  In terms of these variables, the 
Hamiltonian is 
\be
      H = {1\over 2} {4\pi^2\over L^2} l^2.
\ee
The spectrum of $\hat l$ does not depend on the representation chosen, 
and we have again the correct spectrum for the Hamiltonian.   

Notice that we could have written the Hamiltonian (perhaps ``more 
correctly'') as 
\be
      H = {1\over 2} {1\over |g|^2} l^2.
\ee
Then, the corresponding operator is different in each representation, but 
since the representation is tied to $L$, the spectrum is again correct. 
In any case, no free parameter $\iota$ appears.

\subsection{A simple model with the free $\iota$ parameter}

Consider a phase space $\Rl^6$ with coordinates $(K_i,E^i)$ and 
Hamiltonian $H = E_i E_i/2$, and let a theory be defined by 
the constraint $G_i=\epsilon_{ijk} K^j E^k$.  The theory has one 
physical degree of freedom.

Consider the canonical transformation 
\be 
    (K,E) \to (A:=K+\Gamma,E) 
\ee 
where 
\be 
    \Gamma_i:={\partial\sqrt{E^iE^i}\over\partial E_i}.  
\ee 
Here we are mimicking the canonical transformation to the ``connection" 
variable $A$, where $\Gamma$ is the ``spin-connection" of $E$.  In the 
new variables, the constraint becomes 
\be 
     G_i=\epsilon_{ijk}A^j E^k, 
\ee 
which can be viewed as the Gauss law in $d=0$ space dimensions.

Now consider a representation of the canonical commutation relations 
$[\hat{A}_j,\hat{E}^k]=i\delta_j^k$ on the Hilbert space ${\cal 
H}:=L_2(G,d\mu_H(g))$ where $g=\exp(A_i\tau_i)$
is the ``holonomy" of $A$ and $\tau_i$ are the generators of $su(2)$. 

Consider the map $(K,E)\to (K',E') = (\iota K,E/\iota)$.  Under this 
canonical transformation the Hamiltonian scales by $H=E^i 
E^i/2=\iota^2 E'_i E'_i/2$.  The spectrum of the Hamiltonian (Casimir 
of $SU(2)$) is discrete; it is given by $\lambda_n = 
1/2(n/2)[(n/2)+1]$, where $n$ is a positive integer.  But in the 
scaled variables the spectrum of the Hamiltonian changes to $\lambda_n 
= \iota^{2}/2 (n/2)[(n/2)+1]$.  Thus, we have a one parameter 
quantization ambiguity in the theory.

Of course, the choice we have made of the observable algebra for the 
quantization is a bit strange: it is only justified by the analogy 
with gravity.  A more conventional quantization would not lead to any such 
quantization ambiguity.  This shows how the $\iota$ parameter is 
intimately linked with the affine structure of the configuration space 
and with the choice of the holonomies as basic operators in quantum 
gravity.

\section{Conclusions}

In the loop representation approach to quantum gauge theories, 
one does not gauge fix the theory prior to quantization, but rather 
maintains the geometrical structure of the gauge theory explicit in the 
quantum theory.  In particular, one works on the group rather than on the 
algebra. The quantization is based on the physical assumption that 
the Wilson loops, or the ``Faraday lines'' of the theory are the 
physical elementary quantum excitations, and thus correspond to finite norm 
states. Equivalently, the holonomy operator is well-defined.  If there 
is more than one connection ($A$ and $\Gamma$) that can be defined on 
the phase space, and which transform in the same way, then one can 
construct a $\iota$ scaled connection ($A_\iota$) by interpolating 
between distinct connections.  Then, the assumption that the 
elementary physical excitations of the theory are the Wilson loops of 
$A_\iota$ turns out to be physically distinct for different values 
of $\iota$.

This is manifested in the dependence of some physical spectra upon 
$\iota$.  More precisely, what happens in gravity is that the metric 
information is in the conjugate variable $E$.  Being conjugate to the 
connection, $E$ is given in the quantum theory by derivative operators 
acting on functions over the group.  Geometrical quantities which 
are functions of $E$ turn out to be elliptic operators over the group 
manifold.  Hence their discrete spectrum.  But these elliptic 
operators have non-vanishing scaling dimension with respect to the 
affine scaling of the connection.  Therefore, the $\iota$ 
quantization ambiguity shows up in the spectrum of the elliptic 
geometric operators.

The resulting quantization ambiguity, which we have denoted here as 
Immirzi ambiguity, affects the operators with a discrete spectrum that 
scale with $\iota$.  A free parameter $\iota$ appears in the loop 
quantization of general relativity, where it affects the scale of the 
discreteness of space, and in the loop quantization of Maxwell theory, 
where it has been suggested that it might be related to charge 
quantization \cite{k}.  Its origin is not tied to the infinite 
dimensionality of the phase space, nor (at least to a first analysis) 
to the existence of inequivalent representations of the observables 
algebra. 

Finally, the indeterminacy is given by a single parameter, and 
therefore it does not reduce the predictive power of the theory more 
than, say, $\lambda_{qcd}$ in quantum cromodynamics, or the string 
constant in string theory.  Notice in this regard, that also in 
perturbative string theory there are two independent length scales: 
string tension and Planck constant.  Similarly, there are two length 
scales in quantum gravity: the Planck constant $l_{P} = \sqrt{\hbar 
G/c^{3}}$ and the quantum of area $A_{0} = 8 \sqrt{3} \pi \iota 
l_P^2$.  Unless some non yet understood requirement fixes the value of 
the Immirzi parameter, these two length scales are independent.

\section*{Acknowledgments}

We thank Giorgio Immirzi, Jos\`e Antonio Zapata, Alejandro Corichi, Kirill 
Krasnov, Abhay Ashtekar and Don Marolf for useful discussions.  This 
work was supported by DOE-Grant DE-FG02-94ER25228, by NSF Grants 
PHY-5-3840400, PHY-9515506, PHY95-14240 and by the Eberly research fund 
of PSU.

\end{document}